\documentclass{rmaa}
\usepackage{paralist}

\usepackage{psfrag,color}

\SetYear{2004}
\SetConfTitle{-}

\title{WWW database of models of accretion disks irradiated by the central star} 

\author{
  P. D'Alessio,\altaffilmark{1} 
  B. Mer\'{\i}n,\altaffilmark{2}
  N. Calvet,\altaffilmark{3}
  L. Hartmann,\altaffilmark{3}
  and B. Montesinos\altaffilmark{2,4}}

\altaffiltext{1}{Centro de Radioastronom\'\i{}a y Astrof\'{\i}sica, UNAM, Morelia, M\'exico.}
\altaffiltext{2}{Laboratorio de Astrof\'{\i}sica Espacial y F\'{\i}sica Fundamental, INTA, Madrid, Spain.}
\altaffiltext{3}{Harvard-Smithsonian Center for Astrophysics, Cambridge, USA.}
\altaffiltext{4}{Instituto de Astrof\'{\i}sica de Andaluc\'{\i}a, Granada, Spain.} 

\shortauthor{D'Alessio et al.}
\shorttitle{WWW catalog of irradiated accretion disk models}

\fulladdresses{
\item Paola  D'Alessio: Centro de Radioastronom\'\i a y Astrof\'\i sica, 
Universidad Nacional Aut\'onoma de M\'exico, Campus Morelia, 
Apartado Postal 3--72, 58090
  Morelia, Michoac\'an, M\'exico (\email{p.dalessio@astrosmo.unam.mx}).
\item Bruno Mer\'\i n: Laboratorio de Astrof\'{\i}sica Espacial y F\'{\i}sica Fundamental, INTA, Apartado 50727, 28080,  Madrid, Spain (\email{bruno@laeff.esa.es}).
\item Nuria Calvet and Lee Hartmann: Harvard-Smithsonian Center for Astrophysics, 60 Garden Street, Cambridge, MA 02138,  USA. (\email{ncalvet@cfa.harvard.edu}, \email{hartmann@cfa.harvard.edu}).
\item Benjam\'\i n Montesinos: Laboratorio de Astrof\'{\i}sica Espacial y F\'{\i}sica Fundamental, INTA, Apartado 50727, 28080,  Madrid, Spain, and Instituto
de Astrof\'{\i}sica de Andaluc\'{\i}a, CSIC, Camino Bajo de Hu\'etor 24,
18008, Granada, Spain. (\email{bmm@laeff.esa.es}).
}

\abstract{We announce the release of a catalog of physical models of
irradiated accretion disks around young stars based on the modelling
techniques by D'Alessio et al. The WWW catalog includes 
$\sim$ 3000 disk models for different central
stars, disk sizes, inclinations, dust contents and mass accretion
rates. For any of them, radial profiles of disk physical parameters and
synthetic spectral energy distributions can be browsed and downloaded
to compare with observations. 
It can be accessed at {\tt
http://www-cfa.harvard.edu/youngstars/dalessio/} (US), {\tt
http://www.astrosmo.unam.mx/$\sim$dalessio/} (Mexico),
and at {\tt http://www.laeff.esa.es/\-models/dalessio/} (Spain).}

\resumen{Anunciamos la publicaci\'on de un cat\'alogo de modelos
f\'{\i}sicos de discos de acreci\'on irradiados por la estrella
central alrededor de estrellas j\'ovenes, basados en las t\'ecnicas de
modelizaci\'on de D'Alessio et al. El cat\'alogo WWW incluye $\sim$
3000 modelos de disco para diferentes estrellas centrales, tama\~nos
del disco, inclinaciones del disco, contenidos de polvo y tasas de
acreci\'on de masa. Para cada uno de ellos, los perfiles radiales de
las propiedades f\'{\i}sicas del disco y distribuciones espectrales de
energ\'{\i}a sint\'eticas se pueden consultar o descargar desde la
p\'agina web para compararlas con observaciones.
El cat\'alogo est\'a accesible en {\tt http://www-cfa.harvard.edu/youngstars/dalessio/} (EEUU),
{\tt http://www.astrosmo.unam.mx/$\sim$dalessio/}
(M\'exico), y en {\tt http://www.laeff.esa.es/\-models/dalessio/}
(Espa\~na).}

\addkeyword{astronomical database: disk models}
\addkeyword{Stars: Pre-main sequence}
\addkeyword{Stars: Accretion disks}

\begin{document}
% Typeset article header
\maketitle

\section{Introduction}
%The protoplanetary disks where planets may form are thought to be the
%result of the gravitational collapse of the primordial cloud (see
%e.g. Shu, Adams \& Lizano \citeyear{shu87}). 

The old idea that stars are born surrounded by disks, which 
%NC sometimes el may icluye el sometimes, 
may form planetary systems, has found strong observational support in
the last couple of decades.  The properties of these disks are
quantified from the comparison between different observations and
models.  For instance, disk mass accretion rates have been inferred
from the analysis of the short wavelength excess, modeled as produced
by accretion shocks at the stellar surface (Hartigan, Edwards \&
Ghandour 1995; Gullbring et al. 1998; Calvet \& Gullbring 1998;
Hartmann et al. 1998).  Also, mass accretion rates and inner disk
radii have been estimated from models of different line emission
profiles thought to form in the magnetospheric flows connecting the
disks to their central stars (Muzerolle et al. 1998a,b and c;
Muzerolle et al. 2001).  Kinetic information on disks, central star
masses, details of the vertical temperature distribution and molecular
abundances are quantified from detailed analysis of different
molecular lines (e.g., Dutrey et al. 1996; Dutrey, Guilloteau \&
Guelin 1997; Guilloteau \& Dutrey 1998; Simon, Dutrey \& Guilloteau
2000; Najita et al. 2000; Dartois, Dutrey \& Guilloteau 2003; Aikawa
et al. 2002, 2003; Qi et al. 2003; Carr, Tokunaga \& Najita 2004).
Other disk properties, such as masses, degree of flaring of the disk
surface and dust properties, are inferred from the spectral energy
distributions (SEDs) of young stars, from near IR to radio-frequencies,
using different kinds of disk models (e.g., Kenyon \& Hartmann
\citeyear{kh87}; Beckwith et al. 1990; Calvet et al. 1991, 1992;
Malbet \& Bertout 1991; Beckwith \& Sargent 1991; Chiang \& Goldreich
1997, 1999; D'Alessio et al. 1998, 1999, 2001; Dullemond, Dominik \&
Natta (2001); Dullemond, van Zadelhoff \& Natta (2002); Dullemond
(2002); Malbet, Lachaume \& Monin 2001; Lachaume, Malbet \& Monin
2003).  These models go from simple power-law descriptions of the disk
mass surface density, temperature and opacity, to detailed models
where different heating mechanisms are included, and where the
radiative transfer and dust opacity are calculated with different
degrees of sophistication.  Disks have been also  imaged at near IR
(e.g., Stapelfeldt et al. 1998; Koresko 1998; Malbet et al. 1998;
Weinberger et al. 1999; Padgett et al. 1999; Tuthill et al. 2002;
Colavita et al. 2003), mid IR (e.g., McCabe, Duch{\^ e}ne, \& Ghez
2003) and millimeter wavelengths (Dutrey et al. 1996; Wilner, Ho \&
Rodr\'\i guez 1996; Rodr\'\i guez et al. 1998; Mannings \& Sargent 2000).
%producing detailed information about their radial and
%vertical structures 
%NC estas frases no se entienden...
%and, sometimes, revealing inconsistencies of simple
%models in order to fit multiwavelength observations of a given object
%(e.g., Lay, Carlstrom \& Hills 1997; Wilner \& Lay 2000)
%NC querian decir algo asi?
Analysis of these observations with different
models has given information about the radial and
vertical structures of the disks. In general, models
with some degree of complexity are required to
fit multi-wavelength observations of a given object
(e.g., Lay, Carlstrom \& Hills 1997; Wilner \& Lay 2000).

In the present paper we describe a database with a series of 
models of disks around pre-main sequence stars.
These models, which are described in detail in Mer\'\i n 
(2004), are self-consistently calculated  given the properties 
of the central star, the disk 
and its dust, using the methods developed by D'Alessio 
et al. (\citeyear{dalessio98}, \citeyear{dalessio99} 
and \citeyear{dalessio01}). 
We make these models available with the hope that 
they can be useful for fitting  
observations of young stars, helping to  extract more information 
than simpler and usually faster approaches.
%NC
In a forthcoming paper (Mer\'\i n et al. 2005), relationships 
between disk, star and dust parameters and observable quantities such as
spectral indices, colors, etc., will be presented and discussed in detail,
%NC
to enable a faster search for the best model
fit to the observations of a given disk
or a survey of disks.
%NC
In addition,
comparison of models in this database with other modeling 
efforts will be helpful to test the 
underlying assumptions in each set of  models.
%NC
Finally, we hope that this database will help
increase our knowledge of the intrinsic properties
of the disks around young stars.

\section{Disk Models}

The disk models from D'Alessio et al. (\citeyear{dalessio98}, 
\citeyear{dalessio99} and
\citeyear{dalessio01}) have the following characteristics:

\begin{itemize}
%NC
\item Energy is transported by radiation, convection and a turbulent flux, 
the first one being the most important mechanism given the disk parameters 
we have explored in this work.  
\item The disk mass surface density distribution is a consequence of 
%NC
conservation of angular momentum flux given the 
functional form of the viscosity coefficient and the disk 
mass accretion rate.
\item The gas and dust are in thermal balance, 
 and are heated by viscous dissipation, stellar 
irradiation, and ionization by energetic particles 
(from cosmic rays and radioactive decay). 
%NC
For the set of parameters of models in the database,
viscous dissipation is important
in regions close to the star and at the disk midplane, 
%NC
while stellar irradiation is important for most of the disk.
\item The fraction of the stellar radiative flux intercepted by the disk is 
evaluated at the surface where the mean radial optical depth to
the stellar radiation is unity. This surface is self-consistently
calculated given the disk structure and properties of its dust, and 
the intercepted flux is used as a boundary condition in the 
integration of the radiative transfer  equation.
\end{itemize}

%NC
The models are based on the following simplifying assumptions:

\begin{itemize}
\item The disk is in steady state, with a mass accretion \
rate  $ \dot{M} \! = d M \! / \! dt $.
\item The disk is geometrically thin: $H/R\!<<\!1$, where  $H$ is the 
gas scale height 
of the disk and $R$ is the radial distance. 
%NC
With this assumption, 
the vertical and radial structures can be calculated separately.
\item The viscosity coefficient  is given by 
$\nu\!=\!\alpha \, H \, c_s$, following 
the $\alpha$-prescription from Shakura \& Sunyaev (\citeyear{SS73}), where 
$c_s$ is the local sound speed and $\alpha$ is the viscosity parameter, 
assumed to be constant through the disk. The turbulent flux of energy 
is calculated consistently with the $\alpha$ prescription, assuming a 
Prandtl number $P_r=1$.
%NC
\item  Dust and gas are well mixed in the entire disk, 
i.e., the dust to gas mass ratio of each dust ingredient 
and the grains  size distribution are both taken to be constant 
in those regions where the temperature is lower than the sublimation 
temperature.
\item Dust and gas are in thermal equilibrium and a unique 
temperature is calculated for both components.
\item The radiation field is considered in two separate regimes: stellar 
radiation (UV, optical and near IR) and disk radiation 
(IR to radio wavelengths), and 
we use mean opacities calculated with appropriate weighting functions to 
describe 
the interaction of both fields with the disk material.
\item The radiative transfer in each regime is done by solving the two 
first moments of the radiative transfer equation.
\item The inner disk is truncated at the dust destruction radius and the
inner cylindrical surface (``the wall'') emits as a blackbody of 1400 K
(Natta et al. 2001, Muzerolle et al. \citeyear{muzerolle03}). 
%NC
This is a new feature added to the models, which has not 
been included in the models described in the 
previous papers (D'Alessio et al. 1998, 1999, 2001).
%NC como esto es nuevo, hay que decir que se
%hace - no se si es suficiente esto
%BM lo he ampliado un poco para explicar los parametros
% que hemos usado.
We follow the prescriptions in D'Alessio et al. (2003) to calculate
the emission from the wall, taking into account the contributions
of the stellar and accretion luminosities to calculate the position of the
dust destruction radius,  and the dependence on inclination of the
emitting area. More specifically, we compute the inner disk radii with
eq. (3) from Muzerolle et al. (2003) where we take $q \chi_d /
\kappa_d = 2.5$, which gives the best fits to the excess near-IR
continuum in Classical T Tauri disks. Then, we assume a fixed ratio
between the height of the wall and the scale height at the dust
destruction radius, $z_{\rm wall} = 4 H$. However, we list separately the
disk and wall contributions, allowing the user to include different
heights or radii for the wall.
\end{itemize}

%\begin{figure}
%\centering
%\includegraphics[width=0.9\linewidth]{image_temp.ps}
%\caption{Vertical cut of a disk model showing the temperature distribution
%in colors with height and radius (the central star is at the bottom-left).
%In this model the central star has $T_{\rm eff}\!=\!4000\,K$ and $M_*\!=\!2.0\,M_\odot$.
%The disk model has accretion rate $\dot{M}\!=\!10^{-8}\!{\rm M _\odot /yr}$, $R_{\rm disk}\!=\!120\, AU$,
%$\alpha\!=\!0.01$, $p\!=\!3.5$ and $a_{\rm max}\!=\! 1 \mu{\rm m}$.}
%\label{cut}
%\end{figure}

%Fig. \ref{cut} shows and example of the temperature distribution in a disk
%model for a central star with tipical values of $T_{\rm eff}$ and mass of a 
%low-mass pre-main sequence star.

\subsection{Comparison with observations}

The synthetic SED of a model calculated for typical stellar parameters,
mass accretion rate, disk radius, inclination angle and a dust grain size
distribution with millimeter-sized grains, fits the median observed SED of
the Classical T Tauri Stars in the Taurus Molecular Cloud (see D'Alessio et
al. \citeyear {dalessio01}).
 Also, Mer\'{\i}n et al. (2004a) used the models
to fit the detailed SEDs of two HAeBe stars (namely HD 34282 and HD 141569)
and showed that not only the age was responsible for the disk evolution but
also that the metallicity could play an important role.

Allen et al. (2004) have constructed color-color diagrams 
for different clusters using observations from IRAC-Spitzer. 
They find three distinctive regions  in the diagrams: one corresponding 
to stars without IR excess, other  for Classical T Tauri Stars and 
another for embedded objects. The synthetic IRAC colors of 
a subset of models from this library ($T_*=4000$ K and age=1 Myr) 
%NC
agree very well 
with those of observed 
Classical T Tauri Stars (see also Sicilia- Aguilar et al.
2004; Hartmann et al. 2004).

It is important to mention that it is very difficult, if not impossible, 
to constrain the model parameters by 
considering only the SED of a given object. 
The main problem is that the SED is given by  the monochromatic 
flux emerging from 
whole disk, i.e., it is a {\it spatially integrated}
quantity. Thus, detailed information on the spatial distribution 
of the  emergent intensity, which is more directly related to the disk structure,  is hidden in the flux.
Fortunately, different parts of the SED are sensitive 
to different combination of disk parameters, but more than one of these 
combinations can produce the same emergent flux from the disk in 
a particular wavelength range.  
 For instance, the near IR SED is dominated by the 
emission of the wall at the dust sublimation radius. 
For a given sublimation temperature, different silicate  
composition, grain sizes, total (stellar and accretion shock) 
luminosities result in different values of the  sublimation radius, 
but different inclination angles and wall heights can be assumed to 
obtain a similar  wall SED.
In principle, a SED which spans a wide range of wavelengths (from UV to 
radio-frequencies) might help to disentangle some of the disk properties 
when a self-consistent model is used.
For instance, the continuum at mid-IR emerges from optically thick
zones of the disk. In this spectral range, the emergent flux
 depends on the disk mass accretion rate (if the accretion luminosity 
is similar or larger than the
stellar luminosity), the disk inclination angle, and  
the grains composition and size distribution  
(which determines the height of the irradiation surface). The 10 and 18 
$\mu$m silicate bands reflect the silicate composition and grain sizes 
in the atmosphere, thus a self-consistent model should have the same 
type of grains producing the observed features and absorbing the stellar 
radiation which determines the disk vertical temperature distribution and 
its mid-IR emergent continuum.
However, not all the degeneracy can be removed by modeling the SED alone, 
specially if 
the dust at the midplane in the outer disk has a different 
size distribution than the dust in the atmosphere in the inner disk, which is 
expected if dust is settling and growing in disks. 
The only way to remove all possible degeneracies is when the SED and images 
if a self-consistent model are  compared to multi-frequency 
high angular resolution observations.  In this first release of the catalog 
we are making available only SEDs, but intensity distributions at different
wavelengths can be calculated upon request for a given set of model parameters.

%This model used the median accretion rate around low-mass stars of 
%$<\dot{M}>\! = 3\times10^{-8} \! {\rm M_\odot/yr}$ [taken from Gullbring
%et al. \citeyear{gull98} and Hartmann et al. \citeyear{hart98} who 
%estimated this quantity from veiling measurements in young stars].
%The need for millimeter-sized grains to fit the median SED, confirmed
%the idea that some grain growth takes place in protoplanetary disks
%around young stars, from the typical sizes of interstellar grains (sizes
%below 0.25 $\mu$m) to larger particles.

\section{Web-based model library}

We have constructed a grid of models  for different central stars
and a range of values for the physical parameters of the disk
and its dust.

For the central stars, we chose 17 stars with spectral types from K7
to B9 ($T_{\rm eff}\!=\!4000\!-\!10000\, K$ respectively) and ages of
1 and 10 Myrs. Stellar parameters were taken from the pre-main
sequence tracks by Siess et al. (\citeyear{Siessetal00}).

For the disk, we considered four values of the accretion rate, namely
$ \dot{M} \!  = 10^{-6}, 10^{-7}, 10^{-8}, 10^{-9} \, {\rm
M_\odot/y}r$, a typical viscosity parameter
$\alpha\!=\!0.01$, three values of the disk radius $R_{\rm disk}\!=\!
100, \, 300, \, 800 \,$  AU and two inclination angles $i\!=\!30,\, 60$
degrees.

For the dust we use the abundances proposed by Pollack et al. 
 (\citeyear{pollack94}),
a grain size distribution given by $n(a)\!=\!a^{-p}$ with $p\!=\!3.5,\,2.5$,
with a minimum size typical of interstellar grains ($a_{\rm min}\!=\! 0.005
\, \mu{\rm m}$) and six different maximum grain sizes $a_{\rm
max}\!=\!1,\,10,\,100 \, \mu{\rm m},\, 1 \, {\rm mm}, \, 1 \, {\rm
cm}, \, 10 \, {\rm cm}$. 
%NC necesario? ya se describio arriba, aqui se
%estan listando las caracteristicas de los modelos
%It was already shown that the shape of the
%model SEDs depends critically on the dust content (D'Alessio et
%al. \citeyear{dalessio01}).

%NC
%ESTOY DE ACUERDO CON PAOLA QUE UNA TABLA MOSTRANDO
%LAS PROPIEDADES DE LOS MODELOS SERIA MUY ILUSTRATIVA
%PARA EL LECTOR

Columns (1) to (6) of table \ref{cap5tab:input} show the stellar
effective temperatures, spectral types, ages, radii, masses and
luminosities for the central stars in the grid of disk models,
column (7) lists the mass accretion rates of the disks, column (8) 
has the accretion luminosity, and columns (9) and (10) show the 
wall radius, in astronomical units and in stellar radii.
Notice that, for a given central star, the wall radius increases with mass 
accretion rate, reflecting the fact that we are including 
the irradiation from the accretion shocks at the stellar surface as 
an additional heating source of the dust (see Muzerolle et al. 2003).

\begin{table*}
{\small 
\begin{center}
\caption[Parameters for the central stars and disks in the library of disk models]
{Parameters for the central stars and disks in the library of disk
models.}
\vspace{2mm}
  \begin{tabular}{|cccccc|cccc|}
    \hline
\multicolumn{6}{|c}{\bf Central Star} & \multicolumn{4}{|c|}{\bf Irradiated accretion disk model} \\\hline
{\bf T$_{\rm eff}$}  & {\bf  SpType} & {\bf Age}  & ${\bf L_*}$ &  ${\bf R_*}$ & ${\bf M_*}$  & ${\bf \dot{M}}$  & ${\bf L_{\rm acc}}$ &  ${\bf R_{\rm wall}}$ & ${\bf R_{\rm wall}}$ \\
  K            &         &  Myr  &L$_\odot$&R$_\odot$& M$_\odot$& M$_\odot$ yr$^{-1}$ & L$_\odot$& AU & R$_*$ \\\hline
%{\bf  K}            &         & {\bf Myr}  &{\bf L$_\odot$} & {\bf R$_\odot$} & {\bf M$_\odot$} & {\bf M$_\odot$ yr$^{-1}$} & {\bf L$_\odot$} & {\bf AU} & {\bf R$_*$} \\\hline
{\bf 4000}     &{\bf K7} &{\bf  1} &  1.60  & 2.64 &  0.70 &  $1 \times 10^{-9}$ &  0.009 &  0.11 &  8.65 \\
               &         &         &        &      &       &  $1 \times 10^{-8}$ &  0.087 &  0.11 &  8.88 \\
               &         &         &        &      &       &  $1 \times 10^{-7}$ &  0.869 &  0.13 &  10.68 \\
               &         &         &        &      &       &  $1 \times 10^{-6}$ &  8.694 &  0.27 &  21.58 \\\hline
{\bf 4000}     & {\bf K7}&{\bf 10} &  0.31  & 1.16 &  0.80 &  $1 \times 10^{-9}$ &  0.022 &  0.05 &  8.97 \\
               &         &         &        &      &       &  $1 \times 10^{-8}$ &  0.217 &  0.06 &  11.31 \\
               &         &         &        &      &       &  $1 \times 10^{-7}$ &  2.167 &  0.13 &  24.52 \\
               &         &         &        &      &       &  $1 \times 10^{-6}$ &  21.67 &  0.39 &  73.05 \\\hline
{\bf 4500}     &{\bf K4} &{\bf 1}  &  3.76  & 3.20 &  1.40 &  $1 \times 10^{-9}$ &  0.014 &  0.16 &  10.92 \\
               &         &         &        &      &       &  $1 \times 10^{-8}$ &  0.137 &  0.17 &  11.15 \\
               &         &         &        &      &       &  $1 \times 10^{-7}$ &  1.375 &  0.19 &  12.80 \\
               &         &         &        &      &       &  $1 \times 10^{-6}$ &  13.75 &  0.35 &  23.63 \\\hline
{\bf 4500 }    &{\bf K4} &{\bf 3}  & 1.60   & 2.08 &  1.34 &  $1 \times 10^{-9}$ &  0.020 &  0.11 &  11.06 \\
               &         &         &        &      &       &  $1 \times 10^{-8}$ &  0.202 &  0.11 &  11.67 \\
               &         &         &        &      &       &  $1 \times 10^{-7}$ &  2.024 &  0.16 &  16.54 \\
               &         &         &        &      &       &  $1 \times 10^{-6}$ &  20.24 &  0.39 &  40.61 \\\hline
{\bf 4500}     &{\bf K4} &{\bf 10} & 0.60   & 1.28 &  1.10 &  $1 \times 10^{-9}$ &  0.027 &  0.07 &  11.18 \\
               &         &         &        &      &       &  $1 \times 10^{-8}$ &  0.270 &  0.08 &  13.17 \\
               &         &         &        &      &       &  $1 \times 10^{-7}$ &  2.700 &  0.15 &  25.65 \\
               &         &         &        &      &       &  $1 \times 10^{-6}$ &  27.00 &  0.44 &  74.19 \\\hline
{\bf 5000}     &{\bf K1} &{\bf 1}  & 14.86  & 5.14 &  3.00 &  $1 \times 10^{-9}$ &  0.018 &  0.32 &  13.56 \\
               &         &         &        &      &       &  $1 \times 10^{-8}$ &  0.184 &  0.33 &  13.64 \\
               &         &         &        &      &       &  $1 \times 10^{-7}$ &  1.834 &  0.34 &  14.37 \\
               &         &         &        &      &       &  $1 \times 10^{-6}$ &  18.34 &  0.48 &  20.26 \\\hline
{\bf 5000 }    &{\bf K1} &{\bf 10} & 1.72   & 1.75 &  1.40 &  $1 \times 10^{-9}$ &  0.025 &  0.11 &  13.64 \\
               &         &         &        &      &       &  $1 \times 10^{-8}$ &  0.251 &  0.12 &  14.50 \\
               &         &         &        &      &       &  $1 \times 10^{-7}$ &  2.514 &  0.17 &  21.25 \\
               &         &         &        &      &       &  $1 \times 10^{-6}$ &  25.14 &  0.43 &  53.52 \\\hline
{\bf 6000}     &{\bf G0} &{\bf 1}  & 59.10  & 7.13 &  3.50 &  $1 \times 10^{-9}$ &  0.015 &  0.65 &  19.49 \\
               &         &         &        &      &       &  $1 \times 10^{-8}$ &  0.154 &  0.65 &  19.51 \\
               &         &         &        &      &       &  $1 \times 10^{-7}$ &  1.542 &  0.65 &  19.74 \\
               &         &         &        &      &       &  $1 \times 10^{-6}$ &  15.42 &  0.73 &  21.88 \\\hline
{\bf 6000}     &{\bf G0} &{\bf 10} & 5.91   & 2.25 &  1.60 &  $1 \times 10^{-9}$ &  0.020 &  0.20 &  19.56 \\
               &         &         &        &      &       &  $1 \times 10^{-8}$ &  0.223 &  0.21 &  19.89 \\
               &         &         &        &      &       &  $1 \times 10^{-7}$ &  2.234 &  0.24 &  22.92 \\
               &         &         &        &      &       &  $1 \times 10^{-6}$ &  22.34 &  0.45 &  42.70 \\\hline
\end{tabular}

\label{cap5tab:input}
\end{center}
}
\end{table*}

\begin{table*}
\begin{center}
{\small 
\addtocounter{table}{-1}
\caption[(continued)]{(continued).}
\vspace{2mm}
  \begin{tabular}{|cccccc|cccc|}
    \hline
\multicolumn{6}{|c}{\bf Central Star} & \multicolumn{4}{|c|}{\bf Irradiated accretion disk models} \\\hline
{\bf T$_{\rm eff}$}  & {\bf  SpType} & {\bf Age}  & ${\bf L_*}$ &  ${\bf R_*}$ & ${\bf M_*}$  & ${\bf \dot{M}}$  & ${\bf L_{\rm acc}}$ &  ${\bf R_{\rm wall}}$ & ${\bf R_{\rm wall}}$ \\
  K            &         &  Myr  &L$_\odot$&R$_\odot$& M$_\odot$& M$_\odot$ yr$^{-1}$ & L$_\odot$& AU & R$_*$ \\\hline
{\bf 7000}     &{\bf F1} &{\bf 1}  & 130.39 & 7.78 &  4.00 &  $1 \times 10^{-9}$ &  0.016 &  0.96 &  26.53 \\
               &         &         &        &      &       &  $1 \times 10^{-8}$ &  0.161 &  0.96 &  26.54 \\
               &         &         &        &      &       &  $1 \times 10^{-7}$ &  1.615 &  0.97 &  26.69 \\
               &         &         &        &      &       &  $1 \times 10^{-6}$ &  16.15 &  1.02 &  28.12 \\\hline
{\bf 7000}     &{\bf F1} &{\bf 10} & 11.02  & 2.20 &  1.70 &  $1 \times 10^{-9}$ &  0.024 &  0.28 &  27.30 \\
               &         &         &        &      &       &  $1 \times 10^{-8}$ &  0.243 &  0.28 &  27.57 \\
               &         &         &        &      &       &  $1 \times 10^{-7}$ &  2.428 &  0.31 &  30.13 \\
               &         &         &        &      &       &  $1 \times 10^{-6}$ &  24.28 &  0.50 &  48.81 \\\hline
{\bf 8000}     &{\bf A6} &{\bf 1}  & 165.08 & 6.70 &  4.0  &  $1 \times 10^{-9}$ &  0.019 &  1.08 &  34.71 \\
               &         &         &        &      &       &  $1 \times 10^{-8}$ &  0.188 &  1.08 &  34.71 \\
               &         &         &        &      &       &  $1 \times 10^{-7}$ &  1.879 &  1.09 &  34.90 \\
               &         &         &        &      &       &  $1 \times 10^{-6}$ &  18.79 &  1.14 &  36.63 \\\hline
{\bf 8000}     &{\bf A6} &{\bf 10} & 12.62  & 1.85 &  1.9  &  $1 \times 10^{-9}$ &  0.032 &  0.30 &  34.75 \\
               &         &         &        &      &       &  $1 \times 10^{-8}$ &  0.322 &  0.30 &  35.15 \\
               &         &         &        &      &       &  $1 \times 10^{-7}$ &  3.227 &  0.33 &  38.89 \\
               &         &         &        &      &       &  $1 \times 10^{-6}$ &  32.27 &  0.56 &  65.46 \\\hline
{\bf 9000}     &{\bf A2} &{\bf 1}  & 215.39 & 6.05 &  4.0  &  $1 \times 10^{-9}$ &  0.021 &  1.23 &  43.84 \\
               &         &         &        &      &       &  $1 \times 10^{-8}$ &  0.207 &  1.23 &  43.86 \\
               &         &         &        &      &       &  $1 \times 10^{-7}$ &  2.077 &  1.24 &  44.05 \\
               &         &         &        &      &       &  $1 \times 10^{-6}$ &  20.77 &  1.29 &  45.91 \\\hline
{\bf 9000}     &{\bf A2} &{\bf 3}  & 71.00  & 3.47 &  2.7  &  $1 \times 10^{-9}$ &  0.024 &  0.71 &  43.90 \\
               &         &         &        &      &       &  $1 \times 10^{-8}$ &  0.244 &  0.71 &  43.94 \\
               &         &         &        &      &       &  $1 \times 10^{-7}$ &  2.445 &  0.72 &  44.64 \\
               &         &         &        &      &       &  $1 \times 10^{-6}$ &  24.45 &  0.82 &  50.89 \\\hline
{\bf 9000}     &{\bf A2} &{\bf 10} & 17.08  & 1.70 &  2.0  &  $1 \times 10^{-9}$ &  0.037 &  0.34 &  43.99 \\
               &         &         &        &      &       &  $1 \times 10^{-8}$ &  0.370 &  0.35 &  44.41 \\
               &         &         &        &      &       &  $1 \times 10^{-7}$ &  3.697 &  0.38 &  48.46 \\
               &         &         &        &      &       &  $1 \times 10^{-6}$ &  26.97 &  0.62 &  78.16 \\\hline
{\bf 10000}    &{\bf B9.5}&{\bf 1} & 251.94 & 5.30 &  4.0  &  $1 \times 10^{-9}$ &  0.024 &  1.33 &  54.13 \\
               &         &         &        &      &       &  $1 \times 10^{-8}$ &  0.237 &  1.33 &  54.15 \\
               &         &         &        &      &       &  $1 \times 10^{-7}$ &  2.371 &  1.34 &  54.38 \\
               &         &         &        &      &       &  $1 \times 10^{-6}$ &  23.71 &  1.40 &  56.62 \\\hline
{\bf 10000}    &{\bf B9.5}&{\bf 10}& 29.20  & 1.80 &  2.3  &  $1 \times 10^{-9}$ &  0.040 &  0.45 &  54.30 \\
               &         &         &        &      &       &  $1 \times 10^{-8}$ &  0.402 &  0.46 &  54.63 \\
               &         &         &        &      &       &  $1 \times 10^{-7}$ &  4.015 &  0.48 &  57.87 \\
               &         &         &        &      &       &  $1 \times 10^{-6}$ &  40.15 &  0.70 &  83.62 \\
\hline	
\end{tabular}
}
\end{center}
\end{table*}

The disk models were computed in parallel with Linux PCs at the
Harvard-Smithsonian CfA (MA, US), LAEFF (Madrid, Spain) and the Linux
cluster {\tt nostromo} at the CRyA (Morelia, M\'exico) and used a
total CPU time of approximately 8000 hours. 

\section{Concluding remarks}

The WWW catalog contains currently more than 3000 model SEDs and disk
structures. We hope this library of models will be a valuable
tool for analyzing the SEDs of young stars surrounded by accretion 
disks. It is a dynamical website. 
We plan to continue filling the catalog with the emission
maps and new disk models with different degrees of dust
settling to the midplane.

\acknowledgements
We would like to acknowledge the corrections and suggestions
made by an anonymous referee which have helped to improve the 
present manuscript.
This work was supported by NASA through grants AR-09524.01-A from the
Space Telescope Science Institute, and by NASA Origins of Solar
Systems grant NAG5-9670. The numerical calculations were performed on
the linux cluster at CRyA- UNAM, acquired through CONACYT grant
36571-E to Enrique V\'azquez-Semadeni.  PD acknowledges grants from
CONACyT and PAPIIT, DGAPA, UNAM, M\'exico.  B. Mer\'{\i}n wishes to
acknowledge to the INTA for its financial support with a graduate
fellowhip.

\end{document}